\documentstyle[preprint,aps]{revtex}
\begin{document}
\draft
\title{Persistent spins in  the linear diffusion approximation of phase
ordering
 and zeros of stationary gaussian processes}
\author{Bernard Derrida$^{\dag \ddag}$, Vincent Hakim$^{\dag}$
 and Reuven Zeitak$^{\dag}$}
\address{ $\dag$ Laboratoire de Physique Statistique, ENS,
 24 rue Lhomond, 75231 Paris Cedex 05, France.\\
$\ddag$Service de Physique Th\'{e}orique
CE Saclay, F91191 Gif sur Yvette, France.}
\date{\today}
\maketitle
\begin{abstract}
The fraction $r(t)$ of spins which have never flipped up to time $t$ is studied
within a
linear diffusion approximation to phase ordering.
Numerical simulations show that, even in this simple context,
$r(t)$
 decays with time like a power-law  with a non-trival exponent $\theta$
which depends
on the space dimension. The local dynamics at a given  point
is a special case of a  stationary gaussian process of known correlation
function
and the exponent $\theta$ is shown to be
determined by the asymptotic behavior of the
probability distribution of intervals between consecutive zero-crossings of
this process.
An approximate way of computing this distribution is proposed, by taking
 the lengths of the  intervals between successive zero-crossings as
independent
random variables. The approximation gives values of the exponent $\theta$ in
close agreement with the  results of simulations.
\end{abstract}
Phase-ordering and domain growth in systems quenched from a disordered phase
to a two-phase coexistence region has been for a long time a subject of active
research   \cite{revart}.
Recently,
a new facet of this problem has been uncovered
\cite{DBG,Stauf,BDG,krap1,krap2,krap3}. For an Ising model initially in a
random configuration, and evolving according to zero temperature Glauber
dynamics,
 the fraction $r(t)$ of spins which have never flipped up to time $t$
decreases like a power-law \cite{DBG,Stauf}
\begin{equation}
r(t) \sim t^{-\theta} \ \ .
\label{frac1}
\end{equation}
Numerical data indicate that the exponent $\theta$ is non-trivial
 and varies   with
the dimension of space \cite{DBG,Stauf}. Similar exponents have been found
in reaction-diffusion models  with particles
of two different mobilities \cite{krap1,krap3,Cardy}.
The decay (\ref{frac1}) has  also been observed
 for the fraction of surface that has never been wet
in an experiment on the growth of breath figures \cite{beysens}.

The simplicity of Glauber dynamics and its relation to  soluble voter
models has made possible an exact determination of $\theta$ for $1d$
Ising and Potts models \cite{DHP}.
In higher dimensions only
  numerical simulations \cite{Stauf,DOS} or approximate methods \cite{majsi}
have been used so far. A very simple (and rather successful)
description of coarsening for a non-conserved order parameter was proposed
some time ago by Ohta, Jasnow and Kawasaki (OJK) \cite{ojk}. In this OJK
approximation,
the order parameter configuration $\phi(\vec{r},t)$ a time $t$ after the
quench, is the
sign of a gaussian field, initially random, which  evolves according to a
linear
diffusion equation. Namely,
\begin{equation}
\phi(\vec{r},t)= {\mathrm{sign}}[A(\vec{r},t)] \
\end{equation}
with
\begin{equation}
\partial_t A(\vec{r},t) = \nabla^2 A(\vec{r},t) \ \ \ \makebox{and} \ \ \
\langle A(\vec{r},0)A(\vec{r'},0)\rangle = \delta^d(\vec{r}-\vec{r'})
\label{dif}
\end{equation}

In this letter, we first report numerical simulations  showing that even
within this approximation, the fraction $r(t)$ of persistent spins
decays (\ref{frac1}) with a non-trivial $\theta$ which varies with the spatial
dimension $d$.
We then show that determining $r(t)$ for the
OJK dynamics is a particular case of a first-passage problem for correlated
random variables. In a logarithmic time scale,
 these problems can be mapped onto stationary gaussian processes.
The exponent $\theta$ is given by the asymptotic decay of the
 distribution of intervals between zero-crossings for the associated
 process. To our surprise,  given the correlation function of the gaussian
process, the  determination of this asymptotic decay
 turns out to be a
hard unsolved problem \cite{rice,kac,cramer}.
 We propose here a simple approximation which consists in
taking the lengths of successive intervals between zero-crossing as independent
identically distributed random variables. The probability distribution of the
interval lengths is  then chosen so as to reproduce the known sign-correlation
function of the stationary gaussian process\cite{bray2}.

We have  simulated the diffusion equation (\ref{dif}) by using a finite
difference first-order
forward Euler scheme on cubic lattices in  $d=1,2,3$
\begin{equation}
A_i(t+\Delta t)= (1-2 d \Delta t) A_i(t) + \Delta t \sum_{<ij>} A_j(t) \ .
\end{equation}
We used  mostly  $\Delta t=0.1$ on lattices of up to
$10^6$ sites and we measured $r(t)$, the fraction of lattice sites $i$ for
which $A_i$  never
changed sign up to time $t$. Our
results are shown on Fig.1 and indicate that
$r(t)$ decays as in (\ref{frac1}) with $\theta\simeq.12$
 in
$d=1$, $\theta\simeq.18$ in $d=2$ and $\theta\simeq.23$ in $d=3$.
Other choices of $\Delta t$ indicate that
$\theta$ is not affected by the value of $\Delta t$
(as long as $\Delta t<1/2d$ to ensure the stability of the Euler scheme).

The stochastic process at a particular lattice point (the origin, say)
is of the type
\begin{equation}
A(t) = \sum_{x>0} K(x,t) \eta_x
\label{conv}
\end{equation}
where the $\eta_x$ are random gaussian variables
(with $\langle \eta_x^2 \rangle =1$)
and the kernel
$K(x,t)$ is
\begin{equation}
K(x,t)= \makebox{Const} \ \   x^{{d-1 \over 2}}\exp(- x^2/4t)/(4\pi t)^{d/2} \
{}.
\label{Kdif}
\end{equation}
Other examples with different kernels include  usual random walks ($\dot{A}=
\eta_t)$ for which K is the
Heavyside function $K(x,t)= H(t-x)$ and  $\theta=1/2$, and walks
under the influence of a random force ($\ddot{A}=\eta_t$)  for which
 $K(x,t)=(t-x) H(t-x)$ and
 $\theta=1/4$ \cite{mckean,sinai}.  Clearly, the first passage exponent
$\theta$  depends on the {\em shape} of the
kernel $K(x,t)$ and takes differents values for  square, triangular or
diffusive kernels.

It is useful to
note that  (\ref{conv}) is a gaussian
process since it is a linear combination of random gaussian variables.
This process is therefore
entirely characterized by its  pair correlation function $\langle A(t)A(t')
\rangle$. This is
also true of the rescaled variables $X(t)=A(t)/\sqrt{\langle A(t)A(t) \rangle}$
which
can be considered as well since we are only interested in the sign changes
of $A(t)$. In the long time limit, the correlation function
$\langle X(t) X(t') \rangle$ depends only on the time ratio $t/t'$ for the
process considered (\ref{Kdif}).
In a logarithmic time scale, $u=\log(t)$,  if one defines $ Y(u)=X(t)$, $Y(u)$
is therefore a gaussian
stationary process
\begin{equation}
\langle Y(u) Y(u') \rangle = C(u-u')
\end{equation}
 with $C(u)=[1/\cosh(u/2)]^{d/2}$ for the OJK process in dimension $d$
(and $C(u)=\exp(-|u|/2)$ for the usual random walk). $C$ is a rapidly
decreasing function of  $u$. Therefore, $Y(u)$ and $Y(u+l)$
are effectively uncorrelated
for sufficiently large $l$ and it is intuitively clear that
the probability that $Y(u)$ keeps the same sign for an interval of length $l$
decreases exponentially with $l$ for large $l$
\begin{equation}
{\mathrm{Prob}}\{Y(u+l')>0,0<l'<l\} \sim \exp(-\theta l) \ \ \makebox{for} \ \
\ l\gg 1
\label{asyp}
\end{equation}
and $\theta$  measures  the (inverse) decorrelation time
 in the $u$-variable. The asymptotic behavior (\ref{asyp}) implies
that the probability of not flipping in the original $t$ variable from time
$t_1$ to $t_2$ with $t_1 \ll t_2$ is  a power law $(t_1/t_2)^\theta$
as numerically observed above.

In order to compute $\theta$, one needs the large $l$ behavior of  the
distribution
$p(l)$ of the lengths $l$ of  intervals  between consecutive zeros of a
gaussian
stationary process characterized by a given  $C(l)$.
 As observed by Rice more than forty years ago \cite{rice}, this is
a surprisingly hard problem which is  still unsolved
 \cite{kac,mckean,cramer}. So, one has to resort to approximate
methods. Perturbative and variational techniques using the free brownian walk
as a starting point have been  proposed  recently \cite{majsi,vhrz}. However,
the free brownian walk and other similar markovian processes have an
infinite density of zero-crossings and are not a good starting point for cases,
for which $C''(0)$ is finite
like the OJK process,  which have a
finite density of zero-crossings (the density of zeros is given
\cite{rice}
 by $\sqrt{- C''(0)}/\pi$).
We propose here an approximation more suited
to these cases which is inspired
by recent works on the distribution of domain sizes in a  $1d$ coarsening
Ising model \cite{alba} where a similar approximation has been shown to be
rather accurate \cite{bdrz}. It consists in considering the lengths between
consecutive zero-crossings as independent random variables, identically
distributed according to a probability distribution $p(l)$. This
$p(l)$ is determined by
requiring that the sign correlation function is the same for the approximate
independent interval approximation and for
 the original stationary
gaussian process \cite{McF}.

For a  process with correlation function $C(l)$,
the probability $P_{+}(l)$
that $Y(u) Y(u+l) >0$  is given by \cite{rice}
\begin{equation}
P_{+}(l)=\frac{1}{2} +
\frac{1}{\pi} \arcsin(C(l)) \ .
\label{p++g}
\end{equation}
To obtain the corresponding probability for the independent interval process,
it is convenient to determine first the probability $P_{cr}(l)$
that a point at distance $l$ from an up-crossing be positive. Either the
first interval after the up-crossing is longer than $l$,
 or the interval is of size $l_1<l$ and the point is positive with
probability $(1-P_{cr}(l-l_1))$. So $P_{cr}(l)$ satisfies
\begin{equation}
P_{cr}(l)=\int_l^{+\infty} dl_1\  p(l_1) +
\int_0^{l} dl_1 \ p(l_1)\ (1-P_{cr}(l-l_1)) \ .
\end{equation}
Taking Laplace transforms (which we denote
with tilde),
 one obtains
\begin{equation}
\tilde{P}_{cr}(\lambda)=[ \lambda(1+\tilde{p}(\lambda))]^{-1} \ \ .
\label{pcr}
\end{equation}
This can now be used to determine the probability that $Y(u)
Y(u+l) >0$
 for the independent interval process. A given point is at
a distance $l_1$ from the first crossing on its right, with probability
\begin{equation}
Q(l_1)=\frac{\int_{l_1}^{+\infty} p(l)\ dl}{\int_0^{+\infty} l p(l)\ dl}
\,\, \
{\mathrm{or,}}\,\,\,  \ \tilde{Q}(\lambda)=\frac{\tilde{p}(\lambda)-1}{\lambda
\tilde{p}'(0)} \ .
\label{q}
\end{equation}
If  $Y(u)$
and $Y(u+l)$ have the same sign either they belong to the same interval
or $Y(u+l)$ has a sign opposite to points immediately to the right of the
first crossing after $u$, so that
\begin{equation}
P_{+}(l)=\int_l^{+\infty} dl_1 \ Q(l_1)+ \int_0^{l} dl_1\  Q(l_1)
\ [1-P_{cr}(l-l_1)] \ .
\end{equation}
This gives for the Laplace transforms
\begin{equation}
\tilde{P}_{+}(\lambda)= 1/\lambda -\tilde{Q}(\lambda)
\tilde{P}_{cr} (\lambda) \ .
\label{p++}
\end{equation}
Substituting (\ref{pcr}) and (\ref{q}) into (\ref{p++}), one  finally
relates the  distribution $p(l)$ of interval lengths to
the probability $P_+(l)$ that two points at distance $l$
have the same sign for the independent
interval process (or rather their
Laplace transforms)
\begin{equation}
\tilde{p}(\lambda)=\frac{1-[\lambda^2 \tilde{P}_{+}(\lambda)-\lambda]\
\tilde{p}'(0)}{1+[\lambda^2 \tilde{P}_{+}(\lambda)-\lambda]\
\tilde{p}'(0)}
\label{pla}
\end{equation}
For an integrable distribution of interval lengths, $\tilde{p}(\lambda)$
should tend to zero as $\lambda\rightarrow\+\infty$. Using (\ref{pla}),
this gives $\tilde{p}'(0)=1/P'_{+}(0)$ which simply means that the mean
interval size ($-\tilde{p}'(0)$) is the inverse of the density of zeros
($-P'_{+}(0)$).

Equation (\ref{pla}) would be exact if the successive intervals were
uncorrelated. Here, the heart of the approximation
is to choose the exact $P_+(l)$ given by (\ref{p++g}) for a given process and
determine $p(l)$ through (\ref{pla}).
The large $l$  behavior of $p(l)$  is given by the singularity
of its Laplace transform with the largest (negative) real part i.e. by the
largest zero of $P'_{+}(0)=\lambda-\lambda^2 \tilde{P}_{+}(\lambda)$
or, using (\ref{p++g}),
\begin{equation}
\sqrt{-C''(0)}= -{\theta} \int_0^{+\infty} dl \exp(\theta l) \
C'(l) \ [1- C^2(l)]^{-1/2} \ .
\label{eqt}
\end{equation}
In the  OJK case,
$C(l)=[1/\cosh(l/2)]^{d/2}$ . A numerical solution of (\ref{eqt})
gives $\theta\simeq.1203$ for  $d=1$, $\,\theta\simeq.186$ for  $d=2$ and
$\theta\simeq.2358$ for  $d=3$ in very good agreement with the  previous
numerical results. For large $d$
the independent interval approximation
(\ref{eqt}) gives $\theta\sim \kappa \sqrt{d}$ with
$\kappa\simeq.1455$ \cite{no1}. Note that for the case of a free brownian
motion,
$C''(0)= \infty$ and the only possibility to satisfy (\ref{eqt}) is that the
integral on the rhs diverges, leading to $\theta= 1/2$ which is exact.

The whole  distribution  $p(l)$ can also be
obtained by inverting the Laplace transform.
One can obtain the Taylor expansion of $p(l)$ around $l=0$  easily from
(\ref{pla}) and then draw $p(l)$ using Pad\'{e} approximants.
 The obtained shapes for
the diffusive case in $d=1,2,3$ are plotted in Fig.2 and compared with
results of direct simulations of the stationary gaussian processes.

We have shown that non-trivial exponents for the number of  persistent
spins appear even in the  simple OJK approximation of coarsening. Contrarily to
the case of Glauber dynamics, $\theta$ increases with dimension.  This means
the OJK theory  is  too crude an approximation to predict $r(t)$ for Glauber
dynamics.
 It is remarkable that the independent interval approximation gives here such
accurate predictions \cite{alba} for the OJK exponents. Still, it would be
interesting to better understand it, for example by a  maximum entropy
argument. Of course, it would be also interesting to estimate the correlations
$p(l_1,l_2)$  between successive intervals and see  whether one could develop a
systematic method of including these correlations. Another  issue would be to
see under what conditions,  the distributions $p(l)$ obtained via (\ref{pla})
are positive  for general kernels $K$ (as nothing guarantees a priori that
it is so).

Lastly, one could try  to extend the OJK approximation to cases where the order
parameter is more complicated than a scalar (as in the Potts model)
 to see how the above aproximation can be adapted to such cases.

{\bf Acknowledgments} We are grateful to D. Dhar for many enlightening
discussions on the subject of this paper when two of us (BD and VH)
were visiting the Newton Institute (Cambridge, UK). We also thank
A. Bray for informing us of his parallel work
 and J. P. Bouchaud for
 interesting comments.


\begin{figure}
\caption{ Log-log plot of the fraction $r(t)$ of sites $i$ for which $A_i$
has never changed sign {\em vs.} time for the discretized diffusion equation
in dimension $d=1,2,3$. $r(t)$ shows a power-law decay with an exponent
which depends on $d$.}
\label{fig.1}
\end{figure}

\begin{figure}
\caption{The probability distribution function of having an interval
of length $l$ for the gaussian stationary process associated to the
diffusion equation in  $d=1,2,3$ as obtained in the independent
interval approximation (lines) and by direct simulations of the
gaussian processes ($d=1$ circles, $d=2$ plusses, $d=3$ diamonds)
 in Fourier space (1000 realizations of length 3276.8; bin
size 0.1).}
\label{fig.2}
\end{figure}

\begin{thebibliography}{20}
\bibitem{no1} The constant $\kappa$ is solution of the equation
$ 2\sqrt{2}=\kappa\int_0^{+\infty} du \exp(\kappa u) u/\sqrt{\exp(u^2/8)-1}$.
\end{thebibliography}
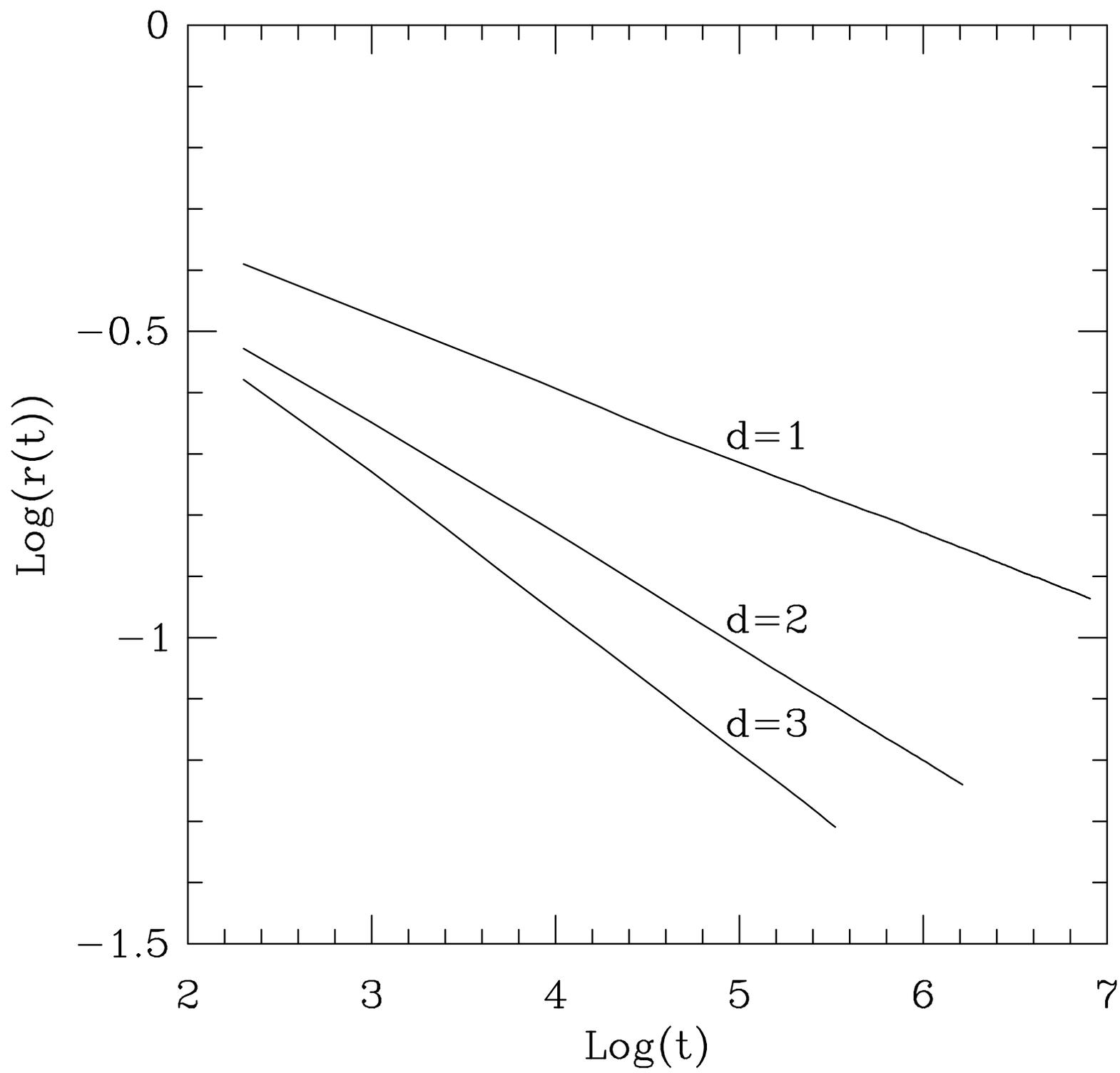
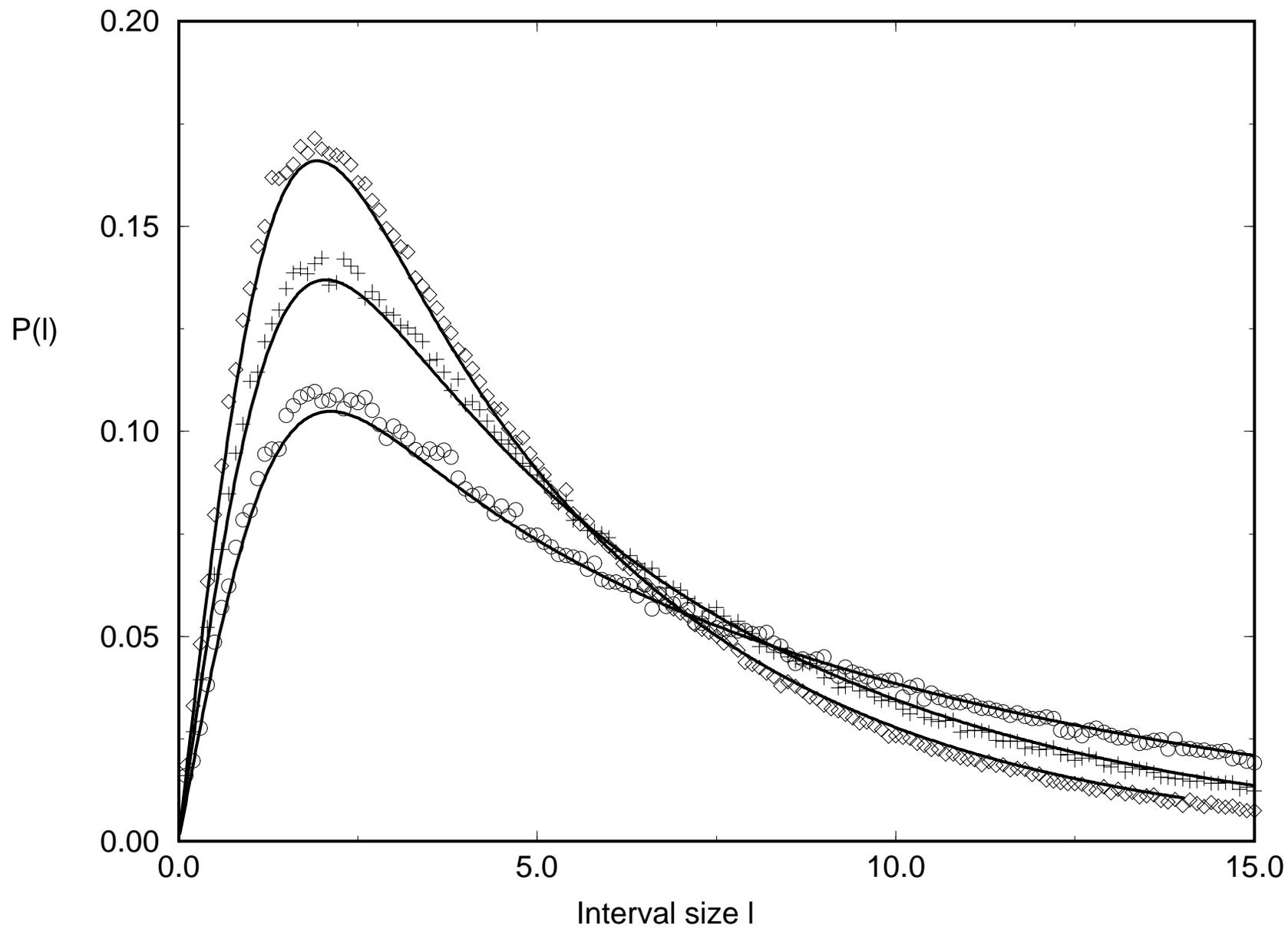
\end{document}